\documentclass[11pt]{article}
\usepackage{arxiv}

\usepackage{amsmath}  
\usepackage{graphicx}
\usepackage{filecontents}
\usepackage{multirow}
\usepackage{xcolor}
\usepackage{soul}
\usepackage{textcomp}
\title{Non-adiabatic transitions and non-equilibrium statistics of deforming nuclei}
\author{
  Nishchal R. Dwivedi$^{*1,2}$, and~Sudhir R. Jain$^{1,3,4}$ \\
  $^1$Nuclear Physics Division, Bhabha Atomic Research Centre, Trombay, Mumbai 400 085, India\\
  $^2$Department of Physics, University of Mumbai, Vidyanagari Campus, Mumbai 400 098, India\\
  $^3$Homi Bhabha National Institute, Training School Complex, Anushakti Nagar, Mumbai 400 094, India \\
  $^4$UM-DAE Centre for Excellence in Basic Sciences, University of Mumbai, Mumbai 400 098, India\\
  $^*$ \texttt{dwivedi.nishchal@gmail.com}
}

\begin{document}
\maketitle
\begin{abstract}
We establish a connection between macroscopic ``heating or cooling" of a finite many-body quantum system and the non-adiabatic Landau-Zener-St\"{u}ckelberg transitions between its quantum states. We have considered the well-known Nilsson model for describing the single-particle states of nuclei and subject the system to a random walk in the deformation space. This subsumes modelling of an evolving many-body system where the dynamics is chaotic. We discover a universality in the distribution of final ``temperatures", beginning with a canonical equilibrium at some temperature $T$. The quantum system is thrown out of equilibrium where free energy and work undergo fluctuations. These fluctuations are shown to respect Jarzynski inequality, and, the Bochkov-Kuzovlev equalities. We believe that this study will pave the way towards understanding non-equlibrium phenomena in other finite quantum systems like metallic clusters, quantum dots, and others.     

\end{abstract}

%\begin{document}
\section{Introduction}
One of the central themes of many-body quantum physics is to relate the relaxation of a system to the evolution of its single-particle energy levels \cite{vanKampen,bob}. Given a many-body system of interacting particles in thermal equilibrium at temperature, $T$, the occupation of the particles is decided by an equilibrium distribution function with a mean-field Hamiltonian \cite{blaizot1986quantum}. There are very well-known constructions describing nuclei and metallic clusters like the Nilsson model \cite{nilsson1955binding} and the Clemenger-Nilsson model \cite{brack1993physics}, billiard models for quantum dots and electromagnetic cavities \cite{jain2017nodal}, and so on. It is also well-known that in this description, when a system evolves in time, the energy levels evolve - they cross, or, avoid to cross, depending on the quantum numbers of their corresponding states \cite{von1993merkwurdige}. For such systems, it has been shown that the energy diffuses in a rigorous manner \cite{jain1999diffusion,jain2000dissipation}. However, for specific case of nuclei, the effect of non-adiabatic transitions on the many-body system, as it is thrown out of canonical equilibrium, has not been understood. It is clear that such a study is broadly relevant to many-body physics, but it is particularly significant as we are engaging in experiments where quantum systems are manipulated in the context of ultracold atomic physics or quantum computing and communication.

In the following, we consider the well-known Nilsson model. This model has played a very important role in understanding various aspects of nuclear structure physics. Choosing this model is also inspired by the general appeal it has to other many-body problems. Moreover, it is possibile to change the shapes and study the evolution of energy levels easily, thus facilitating the investigation we wish to undertake. We extract ``temperature-like" quantity by fitting the Boltzmann distribution to the final state. This is studied for various realizations of shape changes, leading to a distribution of final temperature. These investigations are then fortified by studying fluctuations of observables in the state of non-equilibrium via the well-known works by Jarzynski, Bochkov, Kuzovlev, and Crooks.   
%%%%%%%%%%%%%%%%%%%%%
\section{Nilsson Model}
Here, we consider a complex enough case of nuclei where we are able to understand the nature of state out of equilibrium and see how it emerges as a result of non-adiabatic transitions. We do not claim that our conclusions and results are applicable, in letter, to all many-body systems, but we do believe that they are relevant and useful in spirit. 
The nuclear shell model \cite{mayer1949closed} is successful in accounting for various properties of the nucleus. This approach consists of considering the nucleons moving independently in an averaged potential created by them. By choosing the appropriate form of the potential, this model reproduces the experimentally observed magic numbers. 

In order to understand how the energy levels of a deforming nucleus evolve, Nilsson \cite{nilsson1955binding} presented a model in terms of a three-dimensional oscillator with spin-orbit interaction and a correction to the oscillator potential for higher angular momentum values. This well-known Hamiltonian is given by
\begin{equation}\label{eq:nilsson}
H= H_0 + C \bar{l}.\bar{s} +D \bar{l}^2
\end{equation}
where,
\begin{eqnarray}
H_0=  -\frac{\hbar ^2}{2m} \Delta + \frac{m}{2} (\omega_x^2 x^2+\omega_y^2 y^2+\omega_z^2 z^2).
\end{eqnarray}
Here, $x,y,z$ are the position coordinates of a particle in a frame fixed with respect to the nucleus. $C$ and $D$ are constants which are fixed as discussed in \cite{nilsson1955binding}, $l$ is the orbital quantum number, $s$ is the spin quantum number. For the case of spherical nuclei, $\omega_x=\omega_y=\omega_z=\omega_0=\frac{41}{A^{1/3}\hbar}$. The $\bar{l}^2$ term gives correction to the oscillator potential at large distances and is important for large $l$-values. In order to study how these energy levels change with deformation, we introduce deformation, $\delta$. 

We use:
\begin{eqnarray}
\omega_x^2&=&\omega_0^2\left(1+\frac{2}{3}\delta\right)=\omega_y^2 \nonumber \\
\omega_z^2&=&\omega_0^2\left(1-\frac{4}{3}\delta\right) 
\end{eqnarray}
If we consider that the deformations in the nucleus are volume conserving, then
\begin{equation}
\omega_x\omega_y\omega_z=~\mbox{constant}~= \omega_0^3(\delta=0) = (\omega_0^0)^3
\end{equation}
The above two conditions lead to an expression of $\omega_0$ as a function of deformation as,
\begin{equation}
\omega_0(\delta)= \omega_0^0\left(1-\frac{4}{3}\delta^2 - \frac{16}{27}\delta^3\right)^{-1/6}
\end{equation}
Substitution in the \eqref{eq:nilsson}, modifies the Nilsson Hamiltonian as
\begin{equation}
H = H^0_0 + H_\delta + C ~l\cdot s + D~l^2
\end{equation}
where, $H_\delta = -\delta \hbar \omega_0\frac{4}{3} \sqrt{\frac{\pi}{5}}~r^2~Y_{20}$.

For simplification, we introduce the following parameters,
\begin{eqnarray}
x&=&-\frac{C}{2\hbar\omega^0_0}\nonumber \\
\mu &=& \frac{2D}{C} \nonumber \\
\lambda &=& \frac{\delta}{x}\frac{\omega_0}{\omega^0_0}
\end{eqnarray}
which reduce the Nilsson Hamiltonian \cite{nilsson1955binding} to
\begin{equation}
H=H_0+ x\cdot\hbar\omega^0_0 ~\textrm{R}
\end{equation}
where
\begin{eqnarray}
H_0 &=& \hbar \omega_0 \frac{1}{2} [-\Delta + r^2]   \\
\textrm{R} &=& \lambda~ U - 2~ l\cdot s-\mu l^2 \\
U &=& -\frac{4}{3}\sqrt{\frac{\pi}{5}}r^2 Y_{20}.
\end{eqnarray}
$\lambda$ is the deformation parameter which we can vary to get shapes from prolate to oblate. $\mu$ is a parameter which is varied according to the value of the principal quantum number, $N$. ($\mu=(0,0,0.35,0.45,0.45,0.45,0.40)$ for $N=(1,2,3,4,5,6,7)$, respectively. $x$ is kept a constant i.e. 0.05 by following reference \cite{nilsson1955binding}).

The basis vector chosen is $|Nl\Lambda \Sigma \rangle$ corresponding to the quantum numbers of operators of principal quantum number, $\bar{l}^2$ , $l_z$ and $s_z$. The quantum number $N$ represents the total number of oscillator quanta, such that
\begin{equation}
    H^0_0 |Nl\Lambda \Sigma\rangle = \left(N+\frac{3}{2}\right)\hbar \omega_0 |Nl\Lambda \Sigma\rangle
\end{equation}

The $H^0_0$ term and the $\bar{l}^2$ term of the total Hamiltonian are diagonal in this basis. The matrix elements of the $ \langle l'\Lambda' \Sigma'| l\cdot s |l\Lambda \Sigma\rangle$ term lead to the following selection rules:
\begin{eqnarray}
l&=&l' \nonumber \\
\Lambda &=& \Lambda',~~ \Lambda' \pm 1 \nonumber \\
\Sigma &=& \Sigma' \pm 1,~~ \Sigma'.
\end{eqnarray}

The non-vanishing terms of this matrix elements are given by
\begin{eqnarray}
\langle l,\Lambda \pm1,\mp\frac{1}{2}| l\cdot s |l,\Lambda,\pm\frac{1}{2}\rangle&=&\frac{1}{2}\sqrt{(l\mp\Lambda)(l\pm\Lambda+1)} \nonumber\\~\\
\langle l,\Lambda,\pm\frac{1}{2}| l\cdot s |l,\Lambda,\pm\frac{1}{2}\rangle&=& \pm \frac{1}{2}\Lambda
\end{eqnarray}

The $H_\delta$ term is proportional to the $r^2Y_{20}$ term. This leads to the matrix elements as
\begin{equation}
    \langle l' \Lambda' | Y_{20}| l \Lambda\rangle = \sqrt{\frac{5(2l+1)}{4\pi (2l'+1)}} \langle l 2 \Lambda 0| l 2 l' \Lambda '\rangle \langle 1 2 0 0| l 2 l' 0\rangle
\end{equation}

Thus, upon diagonalisation, this Hamiltonian entails the energy levels which are dependent on deformation in a complicated manner. Due to orthogonal nature of the states, the energy levels of the same spin parity do not intersect. This leads to avoided crossings, where non adiabatic transitions can occur.

Further we define another quantum number $\Omega = \Lambda + \Sigma$, corresponding to the operator $j_z$, which is the total angular momentum operator, which is defined as $j_z = l_z + s_z$. 

The states are characterised as $\Omega^\mathcal{P}$, where $\mathcal{P}=(-1)^{l}$, is the parity given by $(+)$, when $\mathcal{P}=1$ and $(-)$ when $\mathcal{P}=(-1)$

Non-adiabatic transitions have not only been observed in finite systems like metallic clusters \cite{xu2005magnetic} and quantum dots \cite{norris1996measurement} but also their studies have been instrumental in measurement of properties like decoherence for a quantum computer \cite{ashhab2006decoherence}. For studying non-adiabatic transitions in a nucleus, we have considered these `Nilsson' levels, decorating the iconic, so-called Nilsson diagrams.  Non-adiabatic transitions may occur at these avoided crossings whenever the levels come closer than the mean spacing at that deformation \cite{zener1932non}. The ensuing Landau-Zener-St\"{u}ckelberg (LZS) probability between two adjacent levels is given by \cite{zener1932non}
\begin{equation}\label{eq:lzs}
P_{\rm LZS} \sim \exp \left[- \frac{2 \pi}{\hbar}  \frac{(\epsilon _{i}-\epsilon _{i-1})^2}{\bigg| \frac{d(\epsilon _i - \epsilon _{i-1})}{dt}\bigg|}\right] 
\end{equation}
where $\{\epsilon _i\}$ are the single-particle energy levels on the Nilsson diagram. These levels change with deformation, which is time-dependent. Typical time scales, of the order $10^{-21}$ seconds here, correspond to the slowness of the shape changes to preserve the dictates of the adiabatic theorem. However, this breaks down whenever two or more levels come closer than the mean-level spacing at that deformation.
% The denominator can be written as, 
% \begin{eqnarray}
%  \frac{d(\epsilon _i - \epsilon _{i-1})}{dt} &=&  \frac{d(\epsilon _i - \epsilon _{i-1})}{d\lambda}\frac{d\lambda}{dt}
% \end{eqnarray}
 
 We pose the problem of studying how an initially canonical distribution of a nucleus changes as the system is subjected to deformation. We would like to note that there are treatments involving more than two levels which play an important role in the case of complex molecules. 

There is another motivation for carrying out this work, in addition to furthering our understanding of nuclear friction or heating and thermodynamics of such finite systems \cite{beck1995thermodynamics}. 
In a discussion on mass parameters in large amplitude collective motion, a relation was found between microscopic dynamics and diffusive modes for large amplitude collective motion  \cite{jain2012origin}. Mass parameter was shown to originate from to chaotic single-particle motion, and a fractal dimension \cite{foot1} of the ensemble of paths in the deformation space. Thus, it is of great interest to consider a collection of paths in deformation space and let the system evolve; eventually of course we extract average quantities.

Initially, we distribute nucleons among the energy levels according to the canonical distribution for some initial temperature, $T_i$. We then perform a Gaussian random walk of 1000 steps in the deformation space. At each point in the deformation space, the LZS transition probability is calculated and the nucleons are redistributed by these probabilities. At the end of the random walk, a new distribution is obtained, from which a Boltzmann-like temperature is extracted by fitting. We collect 1000 such random walks for various mass numbers ($A$) at 2 and 5 $MeV$. It should be noted that a case with more number of random walks shows similar results, but more number of random walks are computationally expensive. So, an optimal number of 1000 random walks have been chosen.

The total energy is calculated by
\begin{equation}
    E_{total}=\sum_{i=1}^k E_i N_i
    \label{sum}
    \end{equation}
where, $E_i$ is the energy of the $i^{th}$ level at a specific value of deformation and $N_i$ is the number of particles in the $i^{th}$ level. $k$ is the total number of levels between which these nonadiabatic transitions are being studied. Note that this is not an energy-averaged description. That is, we are considering an explicit counting by placing particles on energy levels - $N_i$ for $E_i$, the product is then summed to explain the difference. Let us write equation (\ref{sum}) as

\begin{eqnarray}
E_{total} &=& \int dE \sum^{k}_{i=1} \delta (E-E_i) \nonumber \\
&=& \int dE ~E \sum^{k}_{i=1} N_i \delta(E-E_i)
\label{eq1}
\end{eqnarray}

The sum in equation (\ref{eq1}) is not an averaged quantity, it has fluctuations as $N_i$ depends on the $i^{th}$ state. The total energy in Eq. (\ref{sum}) includes average and fluctuations. If we had taken $N_i \sim N_{avg}$, then Eq. (\ref{eq1}) becomes,

\begin{eqnarray}
E_{Total} &=& \int dE ~E~N_{avg} \sum^{k}_{i=1} \delta(E-E_i)
\label{eq2}
\end{eqnarray}
where the sum in equation (\ref{eq2}) is total level density $g(E)$ which has a decomposition $g_{avg}+\delta g$, where $g_{avg}$ is the average and $\delta g$ is the fluctuating part of the level density. If we had considered $g_{avg}$ alone, we would need to revise the calculations for thermodynamic quantities with correction $\delta g$. 

% The entropy, $S$ is given as,
% \begin{equation}
%     S=p_i \log p_i
% \end{equation}
% where, $p_i=N_i/A$.

\begin{figure}[h!]
\centering
\includegraphics{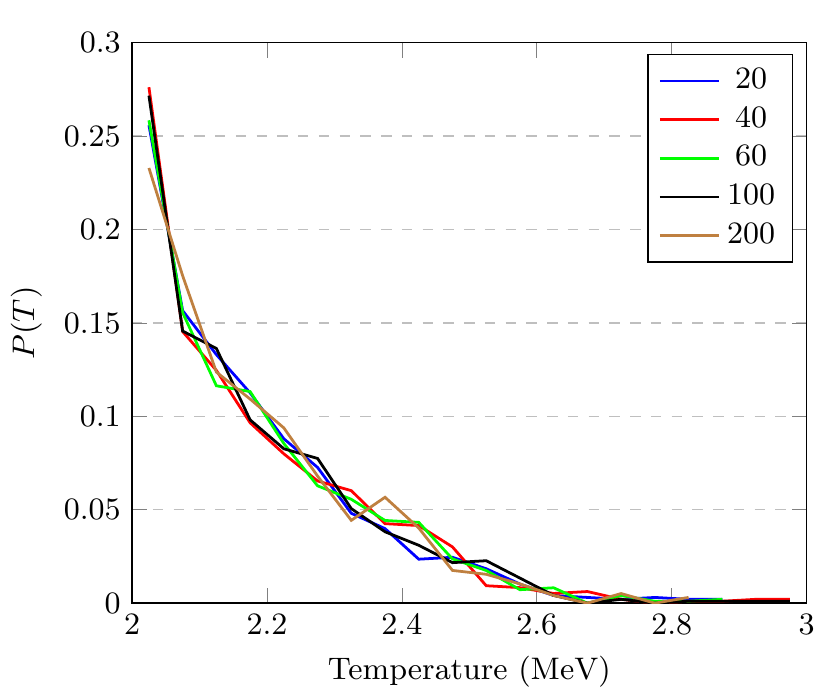}

\caption{An initial Boltzmann distribution for various $A$ values is prepared at $k_b T=2~MeV$ and subjected to a random walk in deformation space. At the end of the deformation, a Boltzmann-like temperature is extracted. Such 1000 realisations are done and are represented as a normalized histogram. The mid points of the histogram are joined to obtain the above figure. It can be seen that the final temperature is always greater than the initial temperature of $2~MeV$.}
\label{fig:2mev}
\end{figure}

\begin{figure}[h!]
\centering
\includegraphics{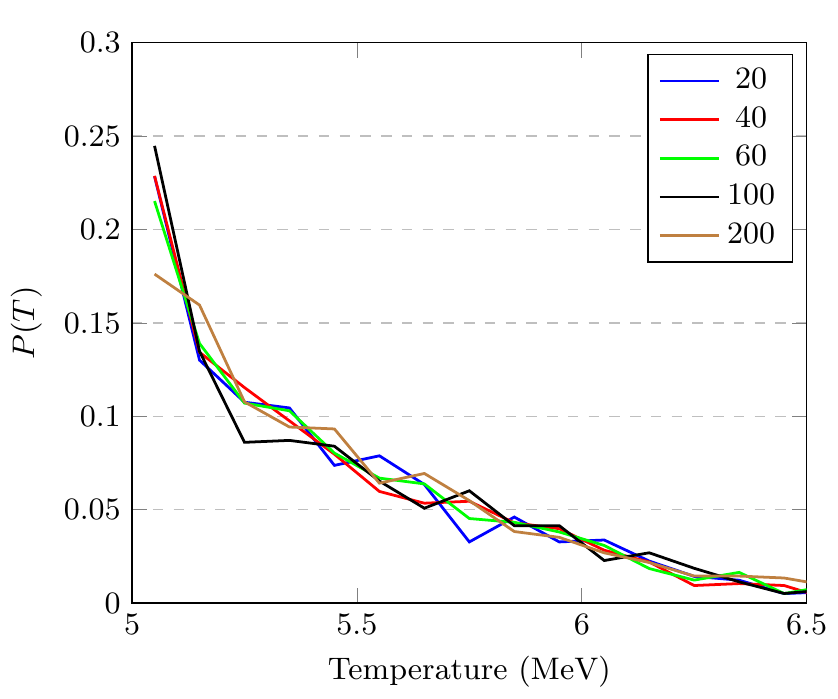}
\caption{An initial Boltzmann distribution for various $A$ values is prepared at $k_b T=5~MeV$ and subjected to a random walk in deformation space. At the end of the deformation, a Boltzmann-like temperature is extracted. Such 1000 realisations are done and are represented as a normalized histogram. The mid points of the histogram are joined to obtain the above figure. It can be seen that the final temperature is always greater than the initial temperature of $5~MeV$.}
\label{fig:5mev}
\end{figure}

Here, we consider the levels in the state $\frac{1}{2}^+$. These levels are populated with $A$ number of particles corresponding to the number of nucleons in the nucleus. We then take the extracted Boltzmann-like ``temperatures" and plot a normalized histogram, which is shown in Figs \ref{fig:2mev} and \ref{fig:5mev}, for $A=20,40,60,100,200$ particles in a nucleus. The $x$-axis denotes the final temperature which the system reaches after the random walk and the $y$-axis denotes the probability distribution function, $P(T)$, of the system to be at that final temperature. Clearly, heating of nuclei is observed, due to the LZS transitions occurring during the deformation of the nucleus.

%2MeV

%5MeV

\section{Universality in temperature trends}

Note that the Fig.\ref{fig:2mev} exhibits a universality in the sense that for a given $T_i$, the trends are similar for all mass numbers, $A$. Same is found for a different $T_i$ in Fig. \ref{fig:5mev}. Thus, for a given $T_i$, there seems to be a universal trend displayed by $P(T)$. 

Furthermore, for different $T_i$'s also, it is possible to show that the trends in Figs \ref{fig:2mev} and \ref{fig:5mev} actually merge into one universal curve. In fact, the distribution of the final temperature depends on the initial $T_i$ as $\exp \left(\frac{-cT}{T_i}\right)$. This is a remarkable observation. 

\begin{figure}[h!]
\centering
\includegraphics{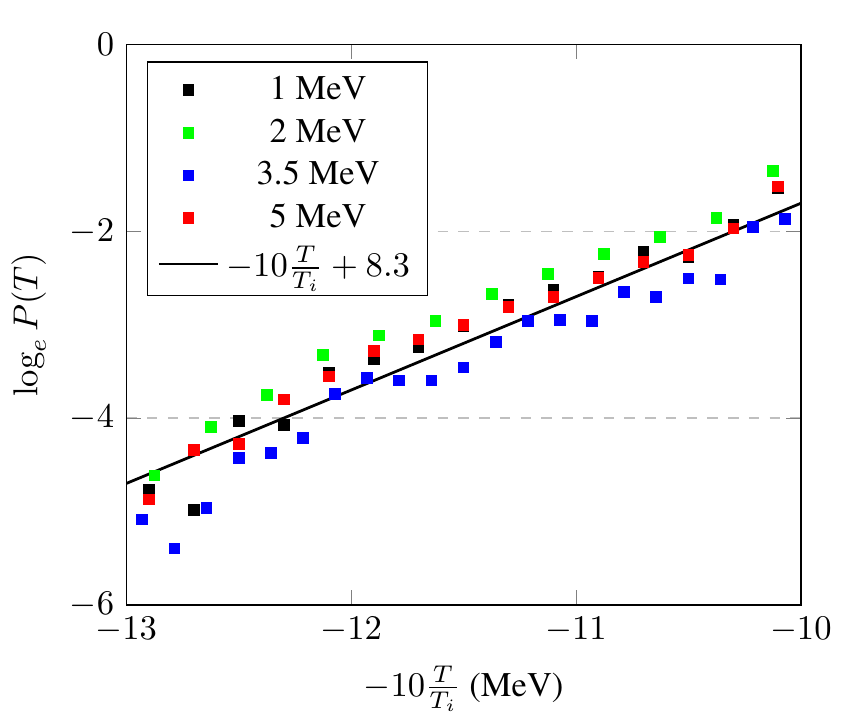}
\caption{We consider the average trends of the LZS heating in nuclei for $k_BT_i=$1, 2, 3.5 and 5 MeV for various $A$ values. Here we plot $\log P(T)$ against $-c\frac{T}{T_i}$, where $c$ is taken to be 10. With this multiplicative factor, we see that the trends of final temperature due to non adiabatic transitions in a system of deforming nuclei, for all initial $T_i$, fall on the line $-c \frac{T}{T_i} + d$ ($c\approx 10$ and $d\approx 8.3$).}
\label{fig:univ}
\end{figure}

To verify this observation, we consider four initial temperatures between 1 MeV and 5 MeV. For each of these temperatures, we extract the average trend for $A= 20, 40, 60, 100, 200$. The average trend is found by evaluating final temperatures by one thousand random walks (of a thousand steps each), for every $A$. The data for each of the $A$ values is then taken together and a normalised histogram is plotted, giving a trend for a given $T_i$. In Fig. \ref{fig:univ} we plot $\log _e P(T)$ versus $- cT/T_i$. We find that the for $c \sim 10$, the points for various $T_i$'s fall on a common curve.  Thus, there seems to be a doubly-universal probability distribution function for the final ``temperatures". This suggest universal law of the heating in the nuclei, originating only from the LZS transitions. 

\section{Jarzynski and Bochkov-Kuzovlev equalities}

%\subsection{Jarzynski equaity}

Consider a thermally isolated system on which external forces deliver work. Then, the Hamiltonian will depend on a time-dependent parameter $\lambda(t)$, which will lead to the evolution of the Hamiltonian. At temperature $T=1/(k_b \beta)$, the average work is given by,
\begin{equation}
    \langle W \rangle = \langle H(p_f, q_f, \lambda(t_f)) - H(p_i, q_i, \lambda (t_i))\rangle
\end{equation}
where the average, $\langle  \rangle$, is performed over repeated experiments from an initial $i$ to final $f$ epochs.

The Jarzynski equality \cite{jarzynski1997nonequilibrium} states that,
\begin{equation}
    \langle \exp{(-\beta W)} \rangle = \exp{(-\beta \Delta F)} 
    \label{jar}
\end{equation}
where, $\Delta F = F_f - F_i$, where $F$ is the equilibrium free energy of the system.

If there is a rapid change in the parameter $\lambda(t)$, then the entropy increases. In such a case,

\begin{equation}
    \langle W \rangle > \Delta F.
    \label{Weq}
\end{equation}

%\subsection{Bochkov - Kuzovlev equality}

For a cyclic process, i.e. when the final and the initial values of the parameter $\lambda(t)$ are the same, then $\Delta F = 0$. In such cases equation \ref{jar} becomes Bochkov - Kuzovlev (BK) equality \cite{bochkov1981-I,bochkov1981-II},
\begin{equation}
    \langle \exp{(- \beta W)} \rangle = 1
    \label{bk}
\end{equation}

%\section{Results}
% We consider the evolution of energy levels due to shape changes in a nucleus in accordance of Nilsson Hamiltonian. Here, the time dependent parameter $\lambda(t)$ corresponds to the change in shape of the nucleus. We consider this smooth shape change governed by a Gaussian random distribution, of 1000 steps. Due to the LZS transitions, the initially distributed particles get redistributed to change the thermodynamics of the system. We calculate the Work done of the system for each random walk. We average  these work to get $\langle W \rangle$. 

To be sure that our study is consistent with fundamental ideas of non-equilibrium statistical mechanics, we now study Jarzynski and Bochkov-Kuzovlev equalities. It is difficult to overstate the significance of these inequalities \cite{pitaevskii2011rigorous}, thus their verification is of importance for validity of our ideas.

$\Delta F$ is given by,
\begin{equation}
    \Delta F = -k_b T \log \langle \exp{-\beta W} \rangle.
\end{equation}

If we write work as $W=\langle W \rangle + \delta W$ , and expand in terms of $\delta W$, we arrive at the expression \cite{hermans1991simple},
\begin{equation}
    \Delta F \approx \langle W \rangle - \frac{\beta \langle (\delta W)^2 \rangle}{2}.
\end{equation}

The values of  $ \langle {\exp{(-\beta W)}} \rangle $ and $  \exp{(-\beta \Delta F)} $ for two temperatures are shown in the Table \ref{tab}. It can be seen that for temperature $2~MeV$, the Jarzynski equality is followed. The Jarzynski equality is also followed for $5~MeV$ till $A=60$. For high temperature and high $A$ values, the inequality (\ref{Weq}) is followed, hinting that due to the increase in entropy Jarzynski equality is not followed.

\begin{table}[h!]
    \centering
    \begin{tabular}{|c| c |c |c| c |}
 \hline
$~$ &  \multicolumn{2}{c|}{$2~$ MeV}  &  \multicolumn{2}{c|}{$5~$ MeV} \\
\hline
$A$ & $ \langle {\exp{(-\beta W)}} \rangle $ & $  \exp{(-\beta \Delta F)} $   &  $ \langle {\exp{(-\beta W)}} \rangle $ & $  \exp{(-\beta \Delta F)} $ \\

 \hline
 \hline
  20 & 1.0001 & 1.0011 &  1.0077 & 1.0117   \\
 40 & 1.0002 & 1.0002 &  1.0513 & 1.1480   \\
 60 & 1.0008 & 1.0083 &  1.1239 & 1.7179   \\
 100 & 1.0040 & 1.0042 &  995.0000  & 50.8406  \\
 200 & 1.0214 & 1.0368 &  989.0000  & $1.01 \times 10^{18}$  \\
 \hline
 \hline
\end{tabular}
    \caption{The Jarzynski equality is tested for various $A$ values for 2 and 5 $MeV$. 1000 instances of a Gaussian random walk with 1000 steps each in the deformation space are considered. The work done and free energy due to LZS transitions are calculated and their values are compared. The Jarzynski equality is valid for all cases except for $5~MeV$ for $A=100,200$. For high excitation and high number of particles, the entropy is high, which leads to the system following equation \ref{Weq} and hence leading in the higher value of $\exp{(-\beta \Delta F)}$.}
    \label{tab}
\end{table}

We construct a set of deformation paths which are governed by Gaussian random distribution and return to the initial condition. This is the case when a deforming nucleus returns to its original shape. In this case (\ref{bk}) is valid, as seen in Table \ref{tab2}.

\begin{table}[h!]
    \centering
    \begin{tabular}{|c| c| c| c| c| }
 \hline
$~$ &  2 $MeV$&  5 $MeV$   \\
\hline
$A$ & $ \langle {\exp{(-\beta W)}} \rangle $ & $ \langle {\exp{(-\beta W)}} \rangle $  \\
 \hline
 \hline
  20 & 1.0000 & 0.9999     \\
 40 & 1.0000 & 0.9999     \\
 60 & 0.9999 & 0.9998     \\
 100 & 0.9998 & 0.9971     \\
 200 & 0.9895 & 0.9176     \\
 \hline
 \hline
\end{tabular}
    \caption{For a cyclic process with the deformations being governed by Gaussian random process, the $\Delta F$ value is equal to zero. For the validity of the BK equality, the value of $ \langle {\exp{(-\beta W)}} \rangle $ should be one. With 1000 random walks of 1000 steps each, this W is calculated for the LZS transitions in the nucleus. The values of the average of $\exp{-\beta W}$ are  tabulated. It can be seen that for cyclic deformation process guided by the Nilsson Hamiltonian, the BK equality is fulfilled.}
    \label{tab2}
\end{table}
%\subsection{Crooks reversal relation}

\begin{figure}[h!]
\centering
\includegraphics{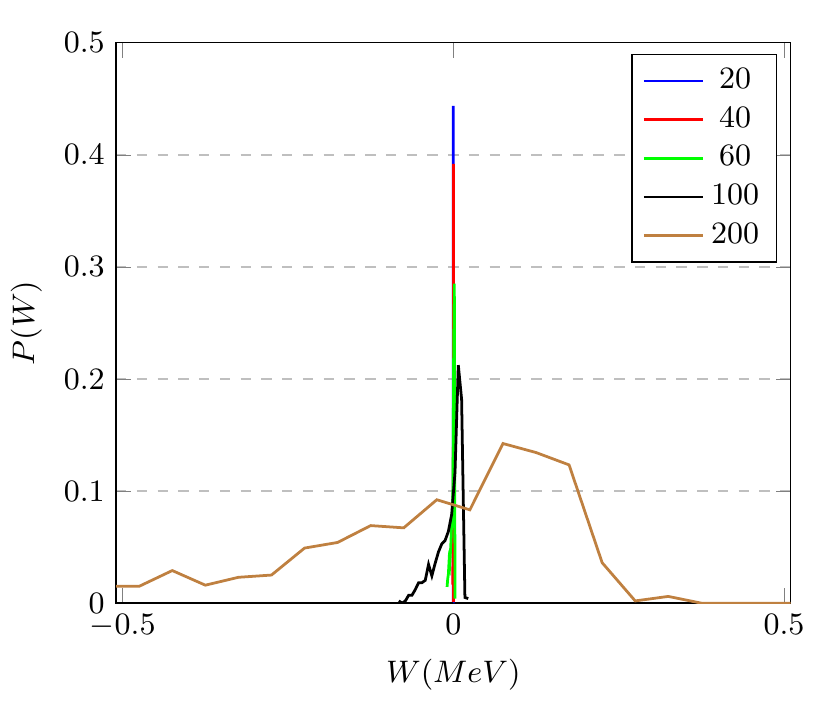}

\caption{The normalized histogram of the distribution of $W$ is shown in the figure above for $2~MeV$ of initial temperature. This histogram is made by collecting 1000 instances of 1000 Gaussian random walks in deformation space each. At the end of each walk, the difference in the initial and final energies gives the work done.}
\label{fig:2mevW}
\end{figure}

\begin{figure}[h!]
\centering
\includegraphics{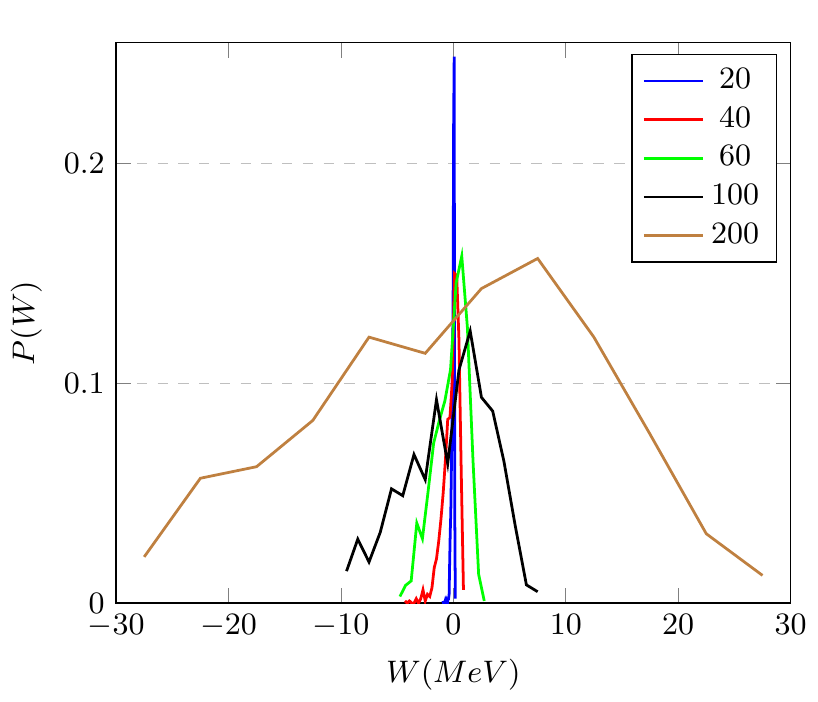}

\caption{The normalized histogram of the distribution of $W$ is shown in the figure above for $5~MeV$ of initial temperature. This histogram is made by collecting 1000 instances of 1000 Gaussian random walks in deformation space each. At the end of each walk, the difference in the initial and final energies gives the work done.}
\label{fig:5mevW}
\end{figure}

Figures \ref{fig:2mevW} and \ref{fig:5mevW} show the probability distribution for work done by LZS transitions along random paths. It can be seen that for heavier nuclei, the distribution of work done is broader as compared to that for lighter nuclei. This is perhaps because in heavier nuclei, the energy levels are closer than that in lighter nuclei. The closeness of the energy levels increases the LZS transitions, leading to more changes in energy of the system and hence the work done. In that it becomes a faster process, thereby verifying (\ref{Weq}).

%Energy

%2MeV

%5MeV

\section{Conclusion}
 We have studied the Landau-Zener-St\"{u}ckelberg transitions among the energy levels of the Nilsson model, describing deforming nuclei of various mass numbers. We have unravelled  a universal behaviour of the probability distribution function of the final temperatures. There is a scale invariance with respect to the temperature, which is quite interesting. The explicit form of the scaling distribution function also helps in quantifying the change in temperature on an average along with the fluctuations. The origin of universality and scale invariance is worth a further investigation. It is possible that the random walks in the deformation space is related to the particular form for $P(T)$.  
 
 The probability density of the (quantum) work done broadens as the number of particles increase due to closely spaced energy levels in heavier nuclei. The verification of the Jarzynski inequality, and, the Bochkov-Kuzovlev equalities for these systems makes the nature of the nonequilibrium state precise. The notion of work done in our context has been carefully considered, following \cite{pitaevskii2011rigorous}.

The collective properties are emerging from a detailed re-organization of occupation of the particles on the energy levels. In usual treatments, these effects are not considered. We believe that self-consistent inclusion of energy level dynamics is crucial, and that this also serves to understand the origin of nonequilibrium phenomena in finite quantum systems. Here, the initial distribution was considered as canonical; however, as shown in \cite{jain1999diffusion}, the results found here are independent of this choice except that the canonical equilibrium helps us to discuss in terms of temperature to start with.

%Heating and dissipation has been a topic of fundamental discussion for a long  time in nuclear physics \cite{jain1998adiabatic}. Here we have shown that one of the mechanisms is the LZS transition by which nuclei evolve to a state which is out of thermodynamic equilibrium. This state is characterized by Jarzynski and Bochkov - Kuzovlev equalities.  

\end{document}